\documentclass[superscriptaddress,twocolumn,preprintnumbers,amsmath,amssymb,aps]{revtex4}

\pdfoutput=1
\usepackage[utf8]{inputenc}

\usepackage{amsmath,amssymb,amsthm,amsfonts}
\usepackage{MnSymbol}
\usepackage{mathrsfs}
\usepackage{slashed}
\usepackage{graphicx}
\usepackage{subfigure}
\usepackage{color,rotating}
\usepackage{dsfont}
\usepackage{setspace}
\usepackage{verbatim}
\usepackage{hyperref,doi}
\usepackage[export]{adjustbox}
\usepackage[normalem]{ulem}
\usepackage{natbib}

\makeatletter
\def\l@subsubsection#1#2{}
\makeatother

\newcommand{\beq}{\begin{equation}}
\newcommand{\eeq}{\end{equation}}
\newcommand{\beqa}{\begin{eqnarray}}
\newcommand{\eeqa}{\end{eqnarray}}
\newcommand{\bfc}{\begin{figure}[t]\begin{center}}
\newcommand{\efc}{\end{center}\end{figure}}

\def\fig#1{Fig.~\ref{#1}}

\def\eq#1{(\ref{#1})}

\def\0#1#2{\frac{#1}{#2}}  

\graphicspath{{./figures/}}

\newcommand{\be}{\begin{eqnarray}}
\newcommand{\ee}{\end{eqnarray}}

\begin{document}

\title{Unsupervised learning of phase transitions: from principal component analysis to variational autoencoders}

\author{Sebastian J. Wetzel} \affiliation{Institut f\"ur Theoretische
  Physik, Universit\"at Heidelberg, Philosophenweg 16, 69120
  Heidelberg, Germany}


\begin{abstract}
 We employ unsupervised machine learning techniques to learn latent parameters which best describe states of the two-dimensional Ising model and the three-dimensional XY model. These methods range from principal component analysis to artificial neural network based variational autoencoders. The states are sampled using a Monte-Carlo simulation above and below the critical temperature. We find that the predicted latent parameters correspond to the known order parameters. The latent representation of the states of the models in question are clustered, which makes it possible to identify phases without prior knowledge of their existence or the underlying Hamiltonian. Furthermore, we find that the reconstruction loss function can be used as a universal identifier for phase transitions. 
\end{abstract}

\maketitle


\section{Introduction}

\noindent Inferring macroscopic properties of physical systems from their microscopic description is an ongoing work in many disciplines of physics, like condensed matter, ultra cold atoms or quantum chromo dynamics. The most drastic changes in the macroscopic properties of a physical system occur at phase transitions, which often involve a symmetry breaking process. The theory of such phase transitions was formulated by Landau as a phenomenological model \cite{Landau1937} and later devised from microscopic principles using the renormalization group \cite{Kadanoff1966,Wilson1975}. One can identify phases by knowledge of an order parameter which is zero in the disordered phase and nonzero in the ordered phase.

Whereas in many known models the order parameter can be determined by symmetry considerations of the underlying Hamiltonian, there are states of matter where such a parameter can only be defined in a complicated non-local way \cite{wen2004}. These systems include topological states like topological insulators, quantum spin hall states \cite{kane2005} or quantum spin liquids \cite{anderson1973}. Therefore, we need to develop new methods to identify parameters capable of describing phase transitions.

Such methods might be borrowed from machine learning. Since the 1990s this field has undergone major changes with the development of more powerful computers and artificial neural networks. It has been shown that such neural networks can approximate every function under mild assumptions \cite{Cybenko1989,Hornik1991}. They quickly found applications in image classification, speech recognition, natural language understanding and predicting from high-dimensional data. Furthermore, they began to outperform other algorithms on these tasks \cite{krizhevsky2012}.

In the last years physicists started to employ machine learning techniques. Most of the tasks were tackled by supervised learning algorithms or with the help of reinforcement learning \cite{curtarolo2003,rupp2012,li2015,ledell2012,pilania2015,saad2012,ovchinnikov2009,arsenault2014,snyder2012,hautier2010,carrasquilla2016,kasieczka2017,carleo2017}. Supervised learning means one is given labeled training data from which the algorithm learns to assign labels to data points. After successful training it can then predict the labels of previously unseen data with high accuracy.

In addition to supervised learning, there are unsupervised learning algorithms which can find structure in unlabeled data. They can also classify data into clusters, which are however unlabelled. It is already possible to employ unsupervised learning techniques to reproduce Monte-Carlo-sampled states of the Ising model \cite{Torlai2016}. Phase transitions were found in an unsupervised manner using principal component analysis \cite{Wang2016,Nieuwenburg2017}. We employ more powerful machine learning algorithms and transition to methods that can handle nonlinear data. A first nonlinear extension is kernel principal component analysis \cite{scholkopf1999}. 

The first versions of autoencoders have been around for decades \cite{bourlard1988,hinton1994} and were primarily used for dimensional reduction of data before feeding it to a machine learning algorithm. They are created from an encoding artificial neural network, which outputs a latent representation of the input data, and a decoding neural network that tries to accurately reconstruct the input data from its latent representation. Very shallow versions of autoencoders can reproduce the results of principal component analysis \cite{Baldi1989}. 

In 2013, variational autoencoders have been developed as one of the most successful unsupervised learning algorithms \cite{kingma2013}. In contrast to traditional autoencoders, variational autoencoders impose restrictions on the distribution of latent variables. They have shown promising results in encoding and reconstructing data in the field of computer vision.

In this work we use unsupervised learning to determine phase transitions without any information about the microscopic theory or the order parameter. We transition from principal component analysis to variational autoencoders, and finally test how the latter handles different physical models. Our algorithms are able to find a low dimensional latent representation of the physical system which coincides with the correct order parameter. The decoder network reconstructs the encoded configuration from its latent representation. We find that the reconstruction is more accurate in the ordered phase, which suggests the use of the reconstruction error as a universal identifier for phase transitions.

Whereas for physicists this work is a promising way to find order parameters of systems where they are hard to identify, computer scientists and machine learning researchers might find an interpretation of the latent parameters.

\section{Models}

\subsection{Ising Model in 2d}

\noindent The Ising model is one of the most-studied and well-understood models in physics. Whereas the one-dimensional Ising model does not possess a phase transition,  the two-dimensional model does.
The Hamiltonian of the Ising model on the square lattice with vanishing external magnetic $h$ field reads
\begin{align}
H=H(\mathbf{S})=-J\sum_{<i,j>_{nn}} s_i s_j \ ,
\end{align}
with uniform interaction strength $J$ and discrete spins $s_i \in \{+1=\ \uparrow,\ -1=\ \downarrow\}$ on each site $i=1,...,N$. The notation $<i,j>_{nn}$ indicates a summation over nearest neighbors. A spin configuration $\mathbf{S}=(s_1,...,s_N)$ is a fixed assignment of a spin to each lattice site, $\Lambda$ denotes the set of all possible configurations $\mathbf{S}$. We set the Boltzmann constant $k_B=1$ and the interaction strength $J=1$ for the ferromagnetic case and $J=-1$ for the antiferromagnetic case. A spin configuration $\mathbf{S}$ can be expressed in matrix form as 
\begin{align}
 \underbar S\ \hat{=} \
\begin{pmatrix}
    \uparrow      & \downarrow& \uparrow & \dots & \uparrow \\
    \downarrow       & \uparrow& \uparrow & \dots & \uparrow\\
   \vdots      & \vdots & \vdots&  & \vdots \\
   \downarrow       & \downarrow & \uparrow & \dots & \downarrow
\end{pmatrix}_{L \times L} \ .
\end{align}
 Lars Onsager solved the two dimensional Ising model in 1944 \cite{onsager1944}. He showed that the critical temperature is $T_c=2/\ln (1+\sqrt{2})=2.269$.

For the purpose of this work, we assume a square lattice with length $L=28$ such that $L \times L=N=784$, and periodic boundary conditions.
We sample the Ising model using a Monte-Carlo algorithm \cite{metropolis1949} at temperatures $ T \in [0,5]$ to generate $50\,000$  samples in the ferromagnetic case and $10\,000$ samples in the antiferromagnetic case.
The Ising model obeys a discrete $\mathbb{Z}_2$-symmetry, which is spontaneously broken below $T_c$. The magnetization of a spin sample is defined as
\begin{align}
M(\mathbf{S})=\frac{1}{N}\sum_i s_i \ .
\end{align}
The partition function
\begin{align}
Z=\sum_{\mathbf{S} \in \Lambda}\exp{(-H(\mathbf{S})/T)}  
\end{align}
allows us to define the corresponding order parameter. It is the expectation value of the absolute value of the magnetization at fixed temperature
\begin{align}
\langle \lVert M(T)\rVert \rangle=\frac{1}{Z}\sum_{\mathbf{S} \in \Lambda}\lVert M(\mathbf{S}) \rVert\exp(-H(\mathbf{S})/T) \ . \label{eq:magnetization}
\end{align}
Similarly, with the help of the matrix $A_{ij}=(-1)^{i+j}$, we define the order parameter, as the expectation value of the staggered magnetization. The latter is calculated from an element-wise product with a matrix form of the spin configurations
\begin{align}
M_{st}=M(\underbar S\odot A) \ . \label{eq:stag_mag}
\end{align}


\subsection{XY Model in 3d}
\noindent The Mermin-Wagner-Hohenberg theorem \cite{mermin1966,hohenberg1967} prohibits continuous phase transitions in $d\leq 2$ dimensions at finite temperature when all interactions are sufficiently short-ranged. Hence, we choose the XY model in three dimensions as a model to probe the ability of a variational autoencoder to classify phases of models with continuous symmetries. The Hamiltonian of the XY model reads

\begin{align}
H(\mathbf{S})=-J\sum_{<i,j>_{nn}} \textbf{s}_i  \cdot \textbf{s}_j \ , 
\end{align}
with spins on the one-sphere $\textbf{s}_i  \in \mathbb{R}^2,  \left\lVert \textbf{s}_i \right\rVert = 1$. Employing $J=1$, the transition temperature of this model is $T_c=2.2017$  \cite{gottlob1993} Using a cubic lattice with $L=14$, such that $N=L^3=2744$, we perform Monte-Carlo simulations to create 10\,000 independent sample spin configurations in the temperature range of $T \in [0,5]$. The order parameter is defined analogously to the Ising model magnetization \eq{eq:magnetization}, but with the $L^2$-norm of a magnetization consisting of two components.


\section{Methods}

\emph{Principal component analysis} \cite{Pearson1901} is an orthogonal linear transformation of the data to an ordered set of variables, sorted by their variance. The first variable, which has the largest variance, is called the first principal component, and so on. The linear function $\langle\cdot,\mathbf{w}\rangle$, which maps a collection of spin samples $(\mathbf{S}_{(1)},...,\mathbf{S}_{(n)})$ to its first principal component, is defined as
\begin{align}
\underset{\Vert \mathbf{w} \Vert = 1}{\arg \max}\,\left[ \sum_j \left((\mathbf{S}_{(j)}-\mathbf{\mu}) \cdot \mathbf{w} \right)^2 \right] \ , \label{eq:pca}
\end{align}
where $\mathbf{\mu}$ is the vector of mean values of each spin averaged over the whole dataset. Further principal components are obtained by subtracting the already calculated principal components and repeating \eq{eq:pca}.

\emph{Kernel principal component analysis} \cite{scholkopf1999} projects the data into a kernel space in which the principal component analysis is then performed. In this work the nonlinearity is induced by a radial basis functions kernel.

\begin{center}
\begin{figure}[b]
\includegraphics[width=0.5\textwidth]{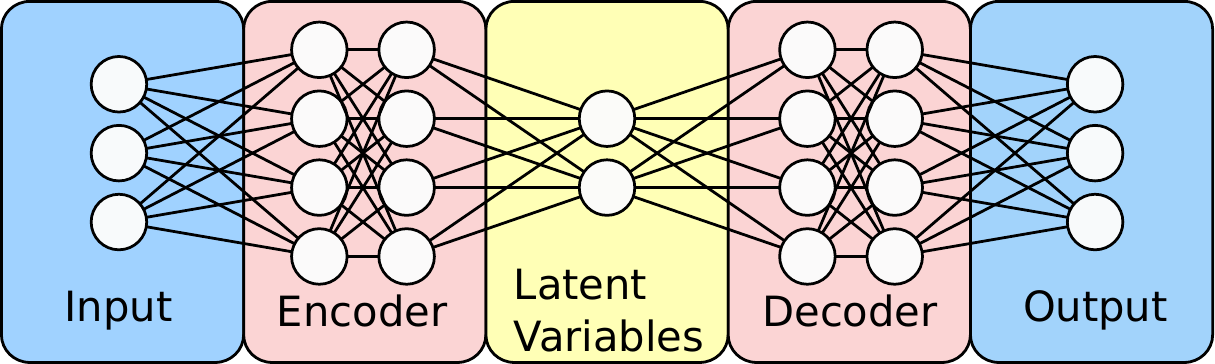}\\
\caption{Neural network architecture}
\label{fig:autoencoder}
\end{figure}
\end{center}

\emph{Traditional neural network-based autoencoders} consist of two artificial neural networks stacked on top of each other. The encoder network is responsible for encoding the input data into some latent variables. The decoder network is used to decode these parameters in order to return an accurate recreation of the input data, shown in \fig{fig:autoencoder}. The parameters of this algorithm are trained by performing gradient descent updates in order to minimize the reconstruction loss (reconstruction error) between input data and output data.

\emph{Variational autoencoders} are a modern version of autoencoders which impose additional constraints on the encoded representations, see latent variables in \fig{fig:autoencoder}. These constraints transform the autoencoder to an algorithm that learns a latent variable model for its input data. Whereas the neural networks of traditional autoencoders learn an arbitrary function to encode and decode the input data, variational autoencoders learn the parameters of a probability distribution modeling the data. After learning the probability distribution, one can sample parameters from it and then let the encoder network generate samples closely resembling the training data.

To achieve this, variational autoencoders employ the assumption that one can sample the input data from a unit Gaussian distribution of latent parameters. The weights of the model are trained by simultaneously optimizing two loss functions, a reconstruction loss and the Kullback-Leibler divergence between the learned latent distribution and a prior unit Gaussian.

In this work we use autoencoders and variational autoencoders \cite{chollet2014} with one fully connected hidden layer in the encoder as well as one fully connected hidden layer in the decoder, each consisting of 256 neurons. The number of latent variables is chosen to match the model from which we sample the input data. The activation functions of the intermediate layers are rectified linear units. The activation function of the final layer is a \emph{sigmoid} in order to predict probabilities of spin $\uparrow$ or $\downarrow$ in the Ising model, or \emph{tanh} for predicting continuous values of spin components in the XY model. We do not employ any $L^1,L^2$ or Dropout regularization. However, we tune the relative weight of the two loss functions of the variational autoencoder to fit the problem at hand. The Kullback-Leibler divergence of the variational autoencoder can be regarded as reguarization of the traditional autoencoder. In our autoencoder the reconstruction loss is the cross-entropy loss between the input and output probability of discrete spins, as in the Ising model. The reconstruction loss is the mean-squared-error  between the input and the output data of continuous spin variables in the XY model.

To understand why a variational autoencoder can be a suitable choice for the task of classifying phases, we recall what happens during training. The weights of the autoencoder learn two things: on the one hand, they learn to encode the similarities of all samples to allow for an efficient reconstruction. On the other hand, they learn a latent distribution of the parameters which encode the most information possible to distinguish between different input samples.

\begin{center}
\begin{figure}[b!]
\includegraphics[width=0.5\textwidth]{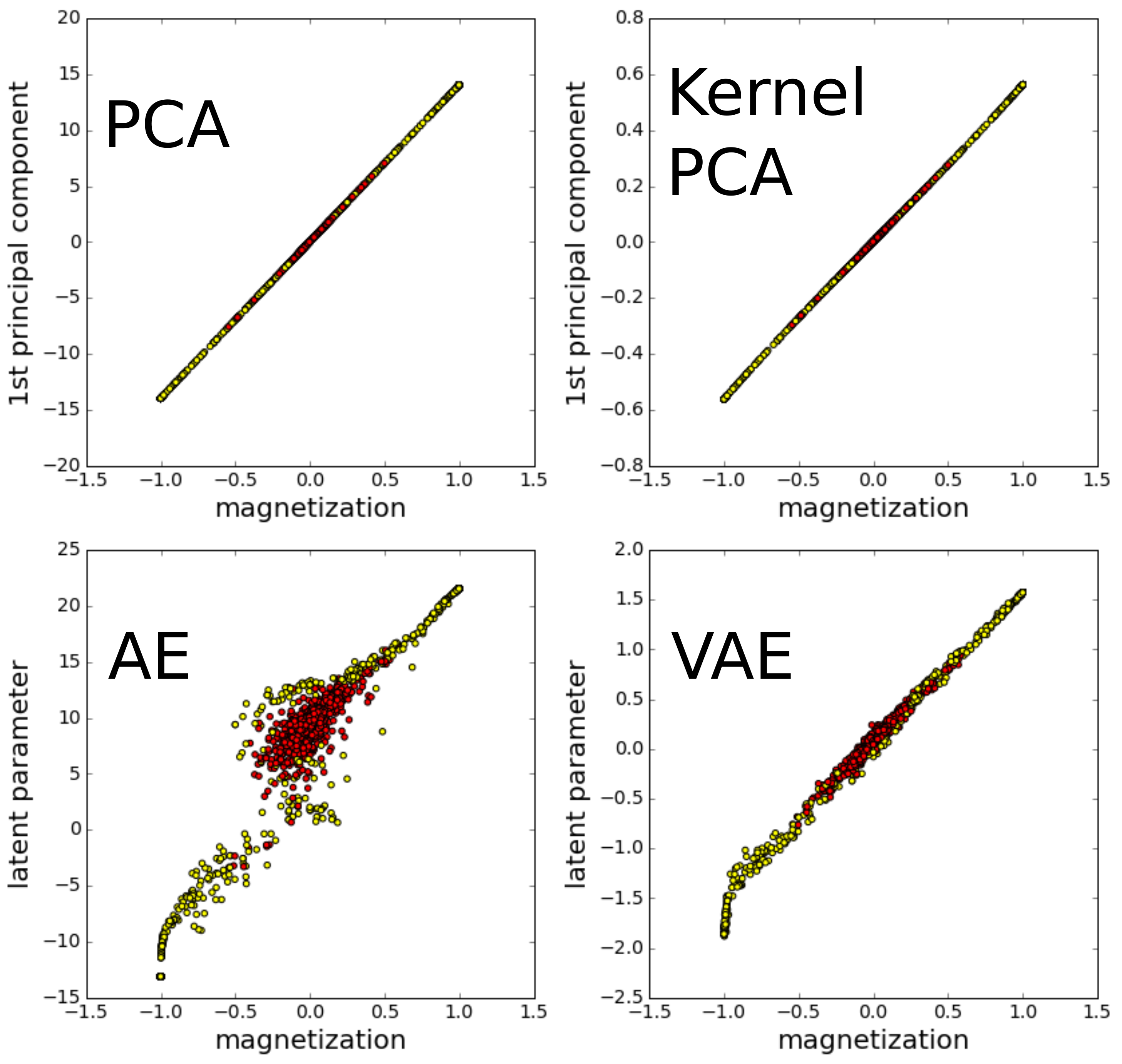}\\
\caption{Ferromagnetic Ising Model. Principal components and latent representations versus magnetization for different algorithms. PCA - principal component analysis, Kernel PCA - kernel principal component analysis, AE - autoencoder, VAE - variational autoencoder.}
\label{fig:compare_methods}
\end{figure}
\end{center}

Let us translate these considerations to the physics of phase transitions. If all the training samples are in the unordered phase, the autoencoder learns the common structure of all samples. The autoencoder fails to learn any random entropy fluctuations, which are averaged out over all data points. However, in the ordered phase there exists a common order in samples belonging into the same phase. This common order translates to a nonzero latent parameter, which encodes correlations on each input sample. It turns out that in our cases this parameter is the order parameter corresponding to the broken symmetry. It is not necessary to find a perfect linear transformation between the order parameter and the latent parameter as it is the case in \fig{fig:compare_methods}. A one-to-one correspondence is sufficient, such that one is able to define a function that maps these parameters onto each other and captures all discontinuities of the derivatives of the order parameter.

We point out similarities between principal component analysis and autoencoders. Although both methods seem very different, they both share common characteristics. Principal component analysis is a dimensionality reduction method which finds the linear projections of the data that maximizes the variance. Reconstructing the input data from its principal components minimizes the mean squared reconstruction error. Although the principal components do not need to follow a Gaussian distribution, principal components have the highest mutual agreement with the data if it emerges from a Gaussian prior. Moreover, a single layer autoencoder with linear activation functions closely resembles principal component analysis \cite{Baldi1989}.
principal component analysis is much easier to apply and in general uses less parameters than autoencoders. However, it scales very badly to a large dataset. Autoencoders based on convolutional layers can reduce the number of parameters. In extreme cases this number can be even less than the parameters of principal component analysis. Furthermore, such autoencoders can promote locality of features in the data.
\begin{center}
\begin{figure*}[t!]
\includegraphics[width=0.9\textwidth]{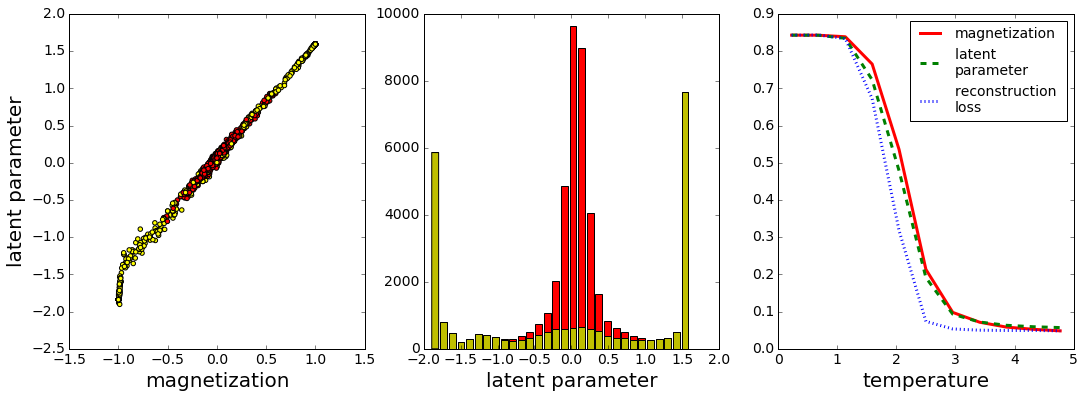}\\
\caption{Ferromagnetic Ising model. Left: correlation between latent parameter and magnetization for each spin sample. Red dots indicate points in the unordered phase, while yellow dots indicate points in the ordered phase. Middle: histogram of occurrences of latent parameters. Red bars correspond to data of the unordered phase, yellow bars correspond to the ordered phase. Right: for each absolute value of magnetization, absolute value of latent parameter and cross-entropy reconstruction loss: average at fixed temperature. The reconstruction loss is mapped on the $T=0$ and $T=5$ value of the magnetization, the latent parameter is rescaled to the magnetization at $T=0$.}
\label{fig:Ising_results}
\end{figure*}
\end{center}
\begin{center}
\begin{figure*}[t!]
\includegraphics[width=0.9\textwidth]{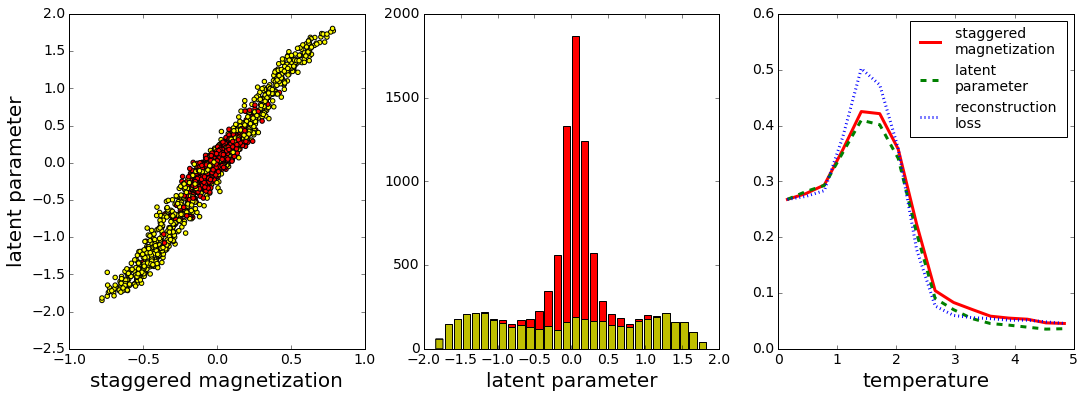}\\
\caption{Antiferromagnetic Ising Model. Left: correlation between latent parameter and staggered magnetization for each spin sample. Red dots indicate points in the unordered phase, while yellow dots indicate points in the ordered phase. Middle: histogram of occurrences of latent parameters. Red bars correspond to data of the unordered phase, yellow bars correspond to the ordered phase. Right: for each absolute value of staggered magnetization, absolute value of latent parameter and cross-entropy reconstruction loss: average at fixed temperature. The reconstruction loss is mapped on the $T=0$ and $T=5$ value of the staggered magnetization, the latent parameter is rescaled to the magnetization at $T=0$.}
\label{fig:IsingAF_results}
\end{figure*}
\end{center}
\begin{center}
\begin{figure}[b!]
\includegraphics[width=0.3\textwidth]{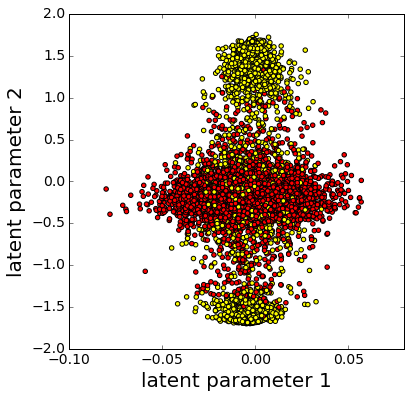}\\
\caption{Ferromagnetic Ising Model. Visualization of data in a two dimensional latent space. Red dots indicate points in the unordered phase, while yellow dots correspond to the ordered phase. The axis for parameter 1 has a smaller range than the axis for parameter 2.}
\label{fig:Ising_latentB}
\end{figure}
\end{center}
\begin{center}
\begin{figure}[bh]
\includegraphics[width=0.5\textwidth]{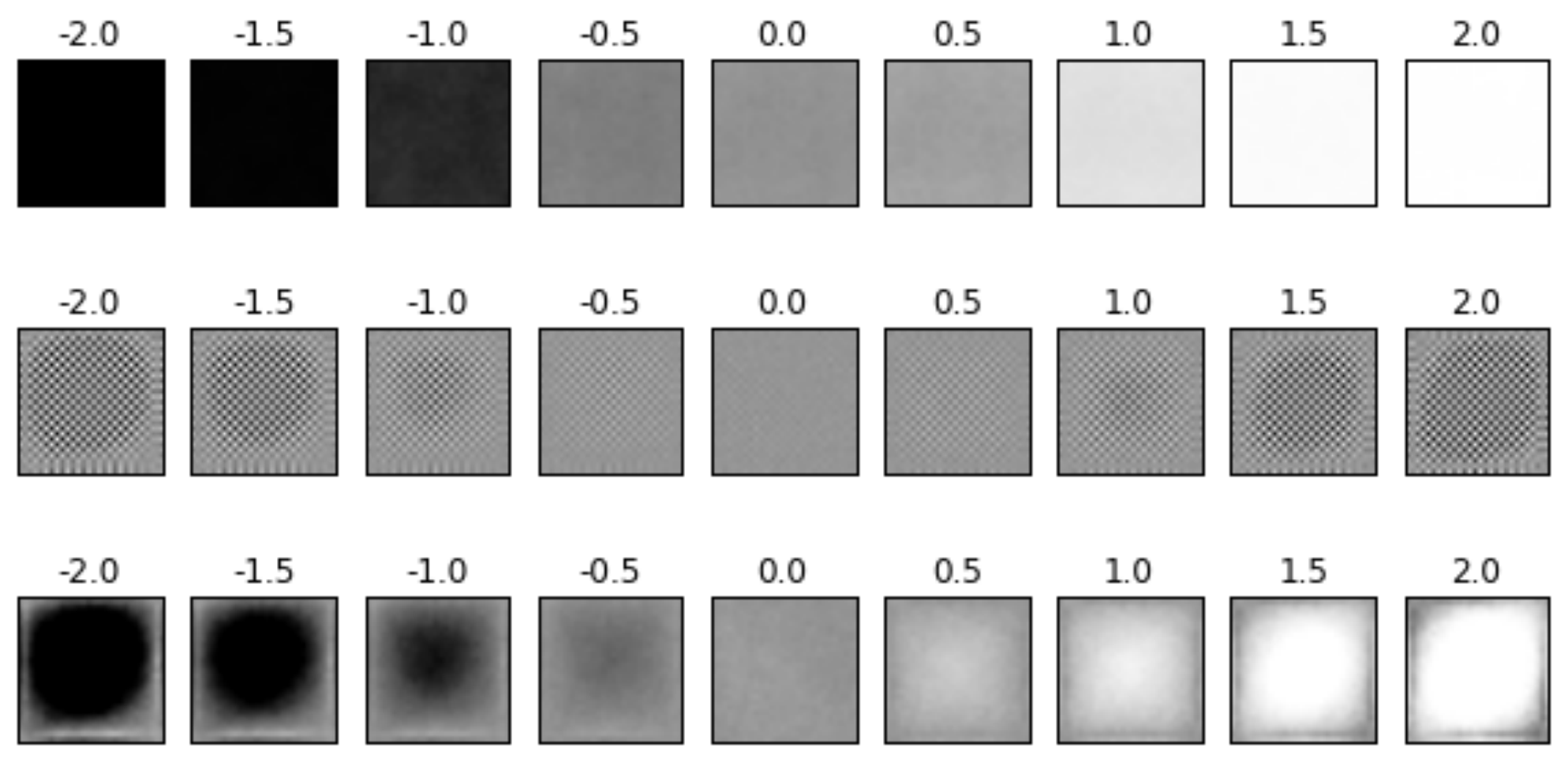}\\
\caption{Reconstruction of images, each consisting of $28\times 28$ pixels, from the latent parameter. The brightness indicates the probability of the spin to be up (white: $p(\uparrow)=1$, black: $p(\downarrow)=1$). The first row is a reconstruction of sample configurations from the ferromagnetic Ising model. The second row corresponds to the antiferromagnetic Ising model. The third row is the prediction from the AF latent parameter, where each second spin is multiplied by $-1$, to show that the second row indeed predicts an antiferromagnetic state.}
\label{fig:reconstruction}
\end{figure}
\end{center}
\begin{center}
\begin{figure*}[t!]
\includegraphics[width=0.9\textwidth]{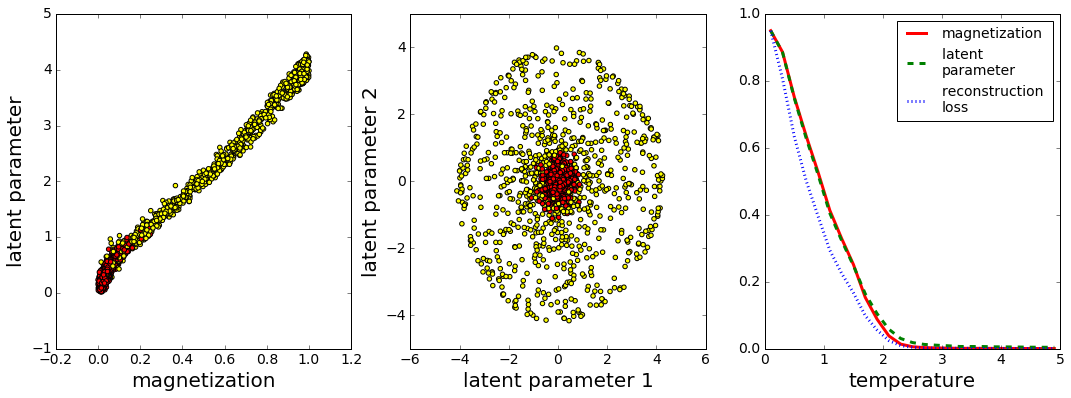}\\
\caption{XY Model. Left: correlation between $L^2$-norm of latent parameter vector and $L^2$-norm of magnetization for each spin sample. Red dots indicate points in the unordered phase, while yellow dots indicate points in the ordered phase. Middle: representation in two-dimensional latent space. Right: for each $L^2$-norm of the magnetization, $L^2$-norm of latent parameter and average of the square root of the mean squared error reconstruction loss: average at fixed temperature. The reconstruction loss is mapped on the $T=0$ and $T=5$ value of the magnetization, the latent parameter is rescaled to the magnetization at $T=0$.}
\label{fig:XY_results}
\end{figure*}
\end{center}
\section{Results}
\subsection{Ising Model}

\noindent The four different algorithms can be applied to the Ising model to determine the role of the first principal components or the latent parameters. \fig{fig:compare_methods} shows a clear correlation between these parameters and the magnetization for all four methods. However, the traditional autoencoder is inaccurate; this fact leads us to enhancing traditional autoencoders to variational autoencoders. The principal component methods show the most accurate results, slightly better than the variational autoencoder. This is to be expected, since the former are modeled by fewer parameters.

In the following results section, we concentrate on the variational autoencoder as the most advanced algorithm for unsupervised learning.

To begin with, we choose the number of latent parameters in the variational autoencoder to be one. After training for 50 epochs and a saturation of the training loss, we visualize the results in \fig{fig:Ising_results}. On the left, we see a close linear correlation between the latent parameter and the magnetization. In the middle we see a histogram of encoded spin configurations into their latent parameter. The model learned to classify the configurations into three clusters. Having identified the latent parameter to be a close approximation to the magnetization $M(S)$ allows us to interpret the properties of the clusters. The right and left clusters in the middle image correspond to an average magnetization of $M(S)\approx \pm 1$, while the middle cluster corresponds to the magnetization $M(S)\approx 0$. Employing a different viewpoint, from \fig{fig:Ising_results} we conclude that the parameter which holds the most information on how to distinguish Ising spin samples is the order parameter. On the right panel, the average of the magnetization, the latent parameter and the reconstruction loss are shown as a function of the temperature. A sudden change in the magnetization at  $T_c\approx2.269$ defines the phase transition between paramegnetism and ferromagnetism. Even without knowing this order parameter, we can now use the results of the autoencoder to infer the position of the phase transition. As an approximate order parameter, the average absolute value of latent parameter also shows a steep change at $T_c$.  The averaged reconstruction loss also changes drastically at $T_c$ during a phase transition. While the latent parameter is different for each physical model, the reconstruction loss can be used as a universal parameter to identify phase transitions. To summarize, without any knowledge of the Ising model and its order parameter, but sample configurations, we can find a good estimation for its order parameter and the occurrence of a phase transition.

It is a priori not clear how to determine the number of latent neurons in the creation of the neural network of the autoencoder. Due to the lack of theoretical groundwork, we find the optimal number by experimenting. If we expand the number of latent dimensions by one, see \fig{fig:Ising_latentB}, the results of our analysis only change slightly. The second parameter contains a lot less information compared to the first, since it stays very close to zero. Hence, for the Ising model, one parameter is sufficient to store most of the information of the latent representation.

While the ferromagnetic Ising model serves as an ideal starting ground, in the next step we are interested in models where different sites in the samples contribute in a different manner to the order parameter. We do this in order to show that our model is even sensitive to structure on the smallest scales. For the magnetization in the ferromagnetic Ising model, all spins contribute with the same weight. In contrast, in the antiferromagnetic Ising model, neighboring spins contribute with opposite weight to the order parameter \eq{eq:stag_mag}.

Again the variational autoencoder manages to capture the traditional order parameter. The staggered magnetization is strongly correlated with the latent parameter, see \fig{fig:IsingAF_results}. The three clusters in the latent representation make it possible to interpret different phases. Furthermore, we see that all three averaged quantities - the magnetization, the latent parameter and the reconstruction loss - can serve as indicators of a phase transition. 

\fig{fig:reconstruction} demonstrates the reconstruction from the latent parameter. In the first row we see the reconstruction from samples of the ferromagnetic Ising model, the latent parameter encodes the whole spin order in the ordered phase. Reconstructions from the antiferromagnetic Ising model are shown in the second and third row. Since the reconstructions clearly show an antiferromagnetic phase, we infer that the autoencoder encodes the spin samples even to the most microscopic level.

\subsection{XY Model}

\noindent In the XY model we examine the capabilities of a variational autoencoder to encode models with continuous symmetries. In models like the Ising model, where discrete symmetries are present, the autoencoder only needs to learn a discrete set, which is often finite, of possible representations of the symmetry broken phase. If a continuous symmetry is broken, there are infinitely many possibilities of how the ordered phase can be realized. Hence, in this section we test the ability of the autoencoder to embed all these different realizations into latent variables.

The variational autoencoder handles this model equally well as the Ising model. We find that two latent parameters model the phase transition best. The latent representation in the middle of \fig{fig:XY_results} shows the distribution of various states around a central cluster. The radial symmetry in this distribution leads to the assumption that a sensible order parameter is constructed from the $L^2$-norm of the latent parameter vector. In \fig{fig:XY_results}, one sees the correlation between the magnetization and the absolute value of the latent parameter vector. Averaging the samples for the same temperature hints to the facts that the latent parameter and the reconstruction loss can serve as an indicator for the phase transition.


\section{Conclusion}
\noindent We have shown that it is possible to observe phase transitions using unsupervised learning. We compared different unsupervised learning algorithms ranging from principal component analysis to variational autoencoders and thereby motivated the need for the upgrade of the traditional autoencoder to a variational autoencoder. The weights and latent parameters of the variational autoencoder approach are able to store information about microscopic and macroscopic properties of the underlying systems. The most distinguished latent parameters coincide with the known order parameters. Furthermore, we have established the reconstruction loss as a new universal indicator for phase transitions. We have expanded the toolbox of unsupervised learning algorithms in physics by powerful methods, most notably the variational autoencoder, which can handle nonlinear features in the data and scale very well to huge datasets. Using these techniques, we expect to predict unseen phases or uncover unknown order parameters, e.g. in quantum spin liquids. We hope to develop deep convolutional autoencoders which have a reduced number of parameters compared to fully connected autoencoders and can also promote locality in feature selection. Furthermore, since there exists a connection between deep neural networks and renormalization group \cite{mehta2014}, it may be helpful to employ deep convolutional autoencoders to further expose this connection.


\emph{Acknowledgments} We would like to thank Timo Milbich, Björn Ommer, Michael Scherer, Manuel Scherzer and Christof Wetterich for useful discussions. We thank Shirin Nkongolo for proofreading the manuscript. S.W. acknowledges support by the Heidelberg Graduate School of Fundamental Physics.


\bibliographystyle{unsrtnat}

\end{document}